# Activation of new Raman modes by inversion symmetry breaking in type II Weyl semimetal candidate $T'$-MoTe$_2$

*Shao-Yu Chen,[†] Thomas Goldstein,[†] Dhandapani Venkataraman,[‡]*

*Ashwin Ramasubramaniam[§] and Jun Yan[†*]*

[†]Department of Physics, University of Massachusetts Amherst, Amherst, MA 01003, USA

[‡]Department of Chemistry, University of Massachusetts Amherst, Amherst, MA 01003, USA,

[§]Department of Mechanical & Industrial Engineering, University of Massachusetts Amherst, Amherst, MA 01003, USA

[*]Corresponding Author: Jun Yan.    Tel: (413)545-0853    Fax: (413)545-1691

E-mail: yan@physics.umass.edu

**Abstract**

We synthesized distorted octahedral ($T'$) molybdenum ditelluride ($MoTe_2$) and investigated its vibrational properties with Raman spectroscopy, density functional theory and symmetry analysis. Compared to the results from high temperature centrosymmetric monoclinic ($T'_{mo}$) phase, four new Raman bands emerge in the low temperature orthorhombic ($T'_{or}$) phase, which was recently predicted to be a type II Weyl semimetal. Crystal-angle-dependent, light-polarization-resolved measurements indicate that all the observed Raman peaks belong to two categories: those vibrating along the zigzag Mo atomic chain ($z$-modes) and those vibrating in the mirror plane ($m$-modes) perpendicular to the zigzag chain. Interestingly the low energy shear $z$-mode and shear $m$-mode, absent from the $T'_{mo}$ spectra, become activated when sample cooling induces a phase transition to the $T'_{or}$ crystal structure. We interpret this observation as a consequence of inversion-symmetry breaking, which is crucial for the existence of Weyl fermions in the layered crystal. Our temperature dependent Raman measurements further show that both the high energy $m$-mode at ~130 cm$^{-1}$ and the low energy shear $m$-mode at ~12 cm$^{-1}$ provide useful gauges for monitoring the broken inversion symmetry in the crystal.

**Keywords**

**Manuscript text**

Distorted octahedral ($T'$) transition metal dichalcogenides (TMDCs) are predicted to possess topologically nontrivial electronic bands that host quantum spin Hall states[1] and type II Weyl fermions[2] in the vicinity of the Fermi energy, which has sparked much recent interest in understanding this class of topological layered compounds. In $T'$-TMDC Weyl semimetals, paired Weyl nodes are located at distinct positions in crystal momentum space as touching points[2,3] between electron and hole pockets, with each node acting as a 'magnetic monopole' emitting Berry flux that leads to anomalous phenomena such as Fermi arcs[4] and violation of chiral charge conservation[5]. While the first $T'$-TMDC proposed to host type II Weyl fermions[2] was $WTe_2$, orthorhombic $T'$-$MoTe_2$ was predicted shortly thereafter to possess similar intriguing band topology with much larger separation between Weyl points of opposite chirality[6], making the Weyl fermions presumably easier to access experimentally with tools such as angle resolved photon emission spectroscopy. This has led to intense experimental studies of $T'$-$MoTe_2$, revealing its rich fundamental properties related to superconductivity, electronic band structure, Fermi surface, lattice vibrations, charge transport, etc.[7-16].

For a nonmagnetic material system, an important condition for the existence of Weyl nodes is the breaking of spatial inversion symmetry. $T'$-$MoTe_2$, as a promising candidate for studying novel type II Weyl physics, has both centrosymmetric and non-centrosymmetric polymorphs, in addition to the hexagonal $H$-$MoTe_2$ that is a semiconductor. In this Letter we take advantage of the sensitivity of Raman spectroscopy to crystal symmetry and probe its lattice vibrations. We observe four new Raman bands in the process of crystalline structural phase transition; in particular, the two shear modes of the crystal, corresponding to the in-plane relative motion of the two MoTe$_2$ atomic layers in the unit cell, appear in the spectra only when inversion symmetry is broken, providing a non-destructive probe that quantitatively gauges symmetry breaking effects in the type II Weyl semimetal candidate. These observations are in interesting contrast to the more widely studied hexagonal *H*-TMDC where the shear mode appears in Raman spectra of the centrosymmetric bulk crystals (see Fig.2(b)), as well as in atomically thin layers either with or without inversion symmetry breaking[17–21].

The polarization-resolved and crystal-angle-resolved Raman measurements performed in this study enable a useful classification of the zone-center lattice vibrations in *T'*-TMDCs in general. Combined with density functional theory (DFT) calculations and symmetry analysis we show that in any monoclinic (*T'$_{mo}$*) or orthorhombic (*T'$_{or}$*), bulk or atomically thin *T'*-TMDC crystal, the zone center phonons can be classified into two types: a third of the modes are *z*-modes that vibrate along the direction of the zigzag transition metal atomic chain (black vertical arrow in Fig.1(d)), while the remaining two-thirds are *m*-modes that vibrate in a mirror plane (red horizontal line in Fig.1(d)) perpendicular to the zigzag chain. The two shear Raman bands we observe belong to *m*- and *z*- mode respectively, and are non-degenerate, reflecting that the *T'*-TMDCs have lower symmetry than their *H* counterparts in which the shear modes are always doubly degenerate[17–21].

The *T'*-MoTe$_2$ crystals used in this work are grown via chemical vapor transport using bromine as the transport agent (details in Methods), as illustrated in Fig.1(a). The as-grown layered crystals have needle-like shape (Fig.1 (b) and (c)), with typical lengths of about 10 mm (along the *a*-axis) and widths of 1 mm (along the *b*-axis). This elongated shape is a result of in-plane anisotropy of the crystal: as illustrated in Fig.1(d) for a monolayer, the strong coupling between the transition metal atoms distorts the crystal lattice, forming zigzag Mo atomic chains (purple zigzags) to lower the free energy of the crystal, resulting in atomic scale periodic buckling. In Fig.1(d) we also illustrate a mirror symmetry plane ($m$, thick red horizontal line) that is perpendicular the zigzag chains. The zigzag ($z$) chain and the mirror ($m$) provide useful classification of the phonons (vibrations along the zigzag chain: $z$-modes; in the mirror plane: $m$-modes) that we will use throughout the paper. Using single crystal X-ray diffraction (XRD), we have determined that the crystal grows preferentially along the zigzag Mo atomic chain ($a$-axis). The room temperature crystal has monoclinic $T'_{mo}$ stacking with a unit cell of $a = 3.493$ Å, $b = 6.358$ Å, $c = 14.207$ Å, $\alpha = \beta = 90°$ and $\gamma = 93.44°$. At low temperatures, the crystal transforms to an orthorhombic phase; a recent study gives the lattice parameters for $T'_{or}$ MoTe$_2$ at 100K as $a = 3.458$ Å, $b = 6.304$ Å, $c = 13.859$ Å, $\alpha = \beta = \gamma = 90°$ [22].

The Raman scattering is performed using crystals with freshly cleaved surface in a cryostat with optical access. A frequency doubled Nd:YAG laser at 532 nm is used to excite the sample. Figure 2(a) schematically shows our experimental setup: the incident light passing through a linear polarizer (LP) is reflected by a beam splitter (BS). A half waveplate is positioned between the BS and the objective lens in order to rotate the polarization of the incident light. Another half waveplate is positioned in the collection path, followed by an LP to select the polarization of interest from the scattered light, which is subsequently dispersed by a triple spectrometer and detected by a liquid-nitrogen-cooled CCD camera. Our measurements are performed in back scattering geometry with incident and scattered light polarized either parallel (HH) or perpendicular (HV) to each other. The direction of incident light polarization with respect to the zigzag Mo chains (crystal $a$-axis) is denoted as angle $\theta$ (Fig.2(a) inset).

Figure 2(b) compares typical room temperature (RT) Raman spectra of distorted octahedral ($T'$) and hexagonal ($H$) MoTe$_2$. The $T'$ crystal displays more Raman bands than $H$, reflecting that the Mo-Mo zigzag atomic chain lowers the crystal symmetry and enlarges the unit cell. Figure 2 (c)-(f) shows detailed RT and LT (low temperature) $T'$ Raman bands with $\theta = 45°$ and $0°$ in HV scattering geometry. The $\theta = 45°$ spectra selectively reveal the $m$-modes (8 for the LT orthorhombic $T'_{or}$ phase and 6 for the RT monoclinic $T'_{mo}$ phase), and the $\theta = 0°$ spectra select the $z$-modes (5 for LT $T'_{or}$ and 3 for RT $T'_{mo}$ phase). Thus in the LT $T'_{or}$ phase, four additional Raman bands become activated as compared with the RT $T'_{mo}$ phase; two of these new bands appear at low energies whereas the other two appear at high energies, as highlighted by the yellow bands in panels (c) and (f). The two new high energy modes are further displayed in the zoomed-in panels (d) and (e), with the spectra being measured by triple additive scattering. As we will show, the low energy $m_{or}$ mode at 12.6 $cm^{-1}$ and $z_{or}$ mode at 29.1 $cm^{-1}$ that appear at LT are the two shear lattice vibrations of the crystal, and their activation directly reflects inversion symmetry breaking in the crystal during the $T'_{mo}$ to $T'_{or}$ phase transition.

For $\theta$ between $0°$ and $45°$, both the $m$- and the $z$-modes have finite intensity. Table 1 displays the detailed Raman intensity dependence on angle $\theta$ for 13 $T'_{or}$ modes and 9 $T'_{mo}$ modes in HH and HV configurations. The angular dependence for $m$-modes in HH scattering is highly sensitive to specific lattice vibration, while in HV all the mode intensities display four-fold symmetry, with the $m$-mode peaks at $\theta = 45°$ and $z$-mode peaks at $\theta = 0°$, as evidenced by spectra in Fig.2 (c) to (f). Below we will explain these experimental observations with symmetry analysis and density functional theory (DFT) calculations.

We have chosen in Fig.1 the in-plane $a$- and $b$- axes as along the zigzag Mo chain and parallel to the mirror plane respectively. The out-of-plane $c$-axis is also parallel to the mirror plane $m$, and is thus perpendicular to the $a$-axis; meanwhile its angle made with the $b$-axis depends on the crystal phase, which is 93.44° in $T'_{mo}$ (Fig.3(b)) and 90° in $T'_{or}$ (Fig.3(c)). The difference in the $c$-axis direction has important consequences for crystal symmetry. To understand this we first examine the symmetry of monolayer $T'$-MoTe$_2$. As illustrated in Fig.3(a), monolayer $T'$-MoTe$_2$ has three symmetry operations in addition to translations along the primitive lattice vectors, including: inversion ($i$), a mirror plane ($m$, see also Fig.1(d)), and a screw axis along the zigzag Mo chain ($2_1^z$, where the superscript $z$ stands for 'zigzag'). These operations, together with the identity operation ($E$), form the $C_{2h}$ group. In bulk crystals, the $T'_{mo}$ phase has the same three symmetries ($i$, $m$, $2_1^z$) as the monolayer. In contrast, the $T'_{or}$ phase only shares the mirror plane symmetry ($m$) with the monolayer. The two other symmetry operations for $T'_{or}$ MoTe$_2$ are a screw axis along $c$-axis ($2_1^c$), and a glide plane perpendicular to the $b$-axis ($n$). The symmetry group of $T'_{mo}$ and $T'_{or}$ MoTe$_2$ are thus $C_{2h}$ (No.11 $P2_1/m$) and $C_{2v}$ (No.31 $Pmn2_1$) respectively[23–25]. For the purpose of discussing the shear modes later, we also illustrated in Fig.3(d) the $H$-MoTe$_2$ unit cell and its inversion centers for comparison. It is worth noting that the inversion centers for $T'_{mo}$-MoTe$_2$ are inside the atomic layer while for $H$-MoTe$_2$, they are in-between the atomic layers.

Since both $T'_{mo}$ and $T'_{or}$ MoTe$_2$ contain two layers of MoTe$_2$ and 12 atoms in the unit cell (shaded area in Fig.3, (b) & (c)), each crystal hosts 36 phonon branches. We use plane-wave density functional theory (DFT) as implemented in the Vienna Ab Initio Simulation Package (VASP)[26] to calculate the 36 phonon branch dispersions. As standard DFT functionals fail to describe interlayer van der Waals bonding correctly, we used the non-local optB86b van der Waals functional[27,28], which reproduces the equilibrium geometry of $MoTe_2$ accurately[29] (see Methods). Tables 2 and 3 show the results of DFT calculation, including first Brillouin zone, phonon dispersion, character table for symmetry group, and zone center normal modes with their calculated energies as well as the vibrational symmetry representations.

We note that all the DFT calculated zone-center optical phonons have different energies; this is because the irreducible representations of both $C_{2h}$ and $C_{2v}$ are one dimensional as shown in the character table in Tables 2 and 3, respectively. This is in contrast to the hexagonal phase, in which in-plane shear, in-plane chalcogen vibrations (IC) and in-plane metal chalcogen vibrations (IMC) are all doubly degenerate[19,21]. The symmetry analysis from Fig.3 also shows that the mirror plane reflection symmetry $m$ is shared by the monolayer, the $T'_{mo}$ and $T'_{or}$ bulk $MoTe_2$ crystals. This provides a generic classification of lattice vibrations in $T'$-TMDC: the vibrations perpendicular to the mirror plane are odd under $m$ ($z$-modes, along the zigzag direction); and the vibrations parallel to the mirror plane are even ($m$-modes). Since motions in the mirror plane (i.e. the $b$-$c$ plane) have twice the number of degrees of freedom as compared with those along the zigzag atomic chain, one third of phonons are odd $z$-modes, and the rest two-thirds are even $m$-modes. To make this clear, we have grouped the 12 $z$-modes and 24 $m$-modes in Tables 2 and 3. We note that this rule can apply to the atomically thin $T'$-TMDC layers as well.

In the back scattering geometry used in our experiment, the Raman active $m$-modes have $A_g$ symmetry in $T'_{mo}$ and $A_1$ symmetry in $T'_{or}$; similarly the Raman active $z$-modes have $B_g$ symmetry in $T'_{mo}$ and $A_2$ symmetry in the $T'_{or}$ phase. The in-plane Raman tensor is thus given by[30]: for the $m$-modes, $\Re_m = \begin{bmatrix} d & 0 \\ 0 & e \end{bmatrix}$ ($A_g$ of $C_{2h}$ for $T'_{mo}$ and $A_1$ of $C_{2v}$ for $T'_{or}$); for the $z$-modes, $\Re_z = \begin{bmatrix} 0 & g \\ g & 0 \end{bmatrix}$ ($B_g$ of $C_{2h}$ for $T'_{mo}$ and $A_2$ of $C_{2v}$ for $T'_{or}$). The intensity of a Raman-active lattice vibration is given by $I = A|\langle \epsilon_i | R^T \cdot \Re \cdot R | \epsilon_o \rangle|^2$, where $A$ is a constant, $\epsilon_i$ and $\epsilon_o$ are polarizations of the incident and outgoing light respectively, $\Re$ is the effective Raman tensor linked to $\Re_m$ or $\Re_z$, $R$ and $R^T$ are the rotation matrix and its transpose that account for rotation of crystal or equivalently, light polarization. The rotation matrix is given by $R = \begin{bmatrix} \cos(\theta) & -\sin(\theta) \\ \sin(\theta) & \cos(\theta) \end{bmatrix}$. In HV scattering, $\epsilon_i = \begin{bmatrix} 1 \\ 0 \end{bmatrix}$, $\epsilon_o = \begin{bmatrix} 0 \\ 1 \end{bmatrix}$.

Thus the Raman intensities are given by:

$$I_{HV}^m(\theta) = \left| \frac{d-e}{2} \right|^2 \sin^2(2\theta),$$

$$I_{HV}^z(\theta) = |g|^2 \cos^2(2\theta). \qquad (1)$$

In HH scattering, $\epsilon_i = \begin{bmatrix} 1 \\ 0 \end{bmatrix}$, $\epsilon_o = \begin{bmatrix} 1 \\ 0 \end{bmatrix}$. Thus the Raman intensities are given by:

$$I_{HH}^m(\theta) = |d \cos^2\theta + e \sin^2\theta|^2,$$

$$I_{HH}^z(\theta) = |g|^2 \sin^2(2\theta). \qquad (2)$$

In HV scattering, the intensities of $m$ and $z$ modes are expected to depend on $\theta$ as $\sin^2(2\theta)$ and $\cos^2(2\theta)$, providing convenient classification of the two types of Raman bands. We have taken advantage of this fact in Fig.2(c)-(f) to selectively display $m$-modes with HV $\theta = 45°$, and $z$-modes with HV $\theta = 0°$. We note that due to in plane anisotropy in dielectric constant and absorption, $d$, $e$, and $g$ in the effective Raman tensors are allowed to be complex[31−33]. Thus in HH scattering, the $z$-mode scales as $\sin^2(2\theta)$ while the angular dependence of the $m$-mode is sensitive to the phase difference between $d$ and $e$, and can exhibit different shapes for different phonons with the same symmetry. These are in good agreement with the angular patterns seen in Table 1 and support our experimental classification and assignment of $m$ and $z$ mode vibrations. With this thorough understanding of symmetry representation and lattice classification, we can unambiguously assign $m_{or}$ at 12.6 cm⁻¹ as the shear mode vibrating along the $b$-axis, and $z_{or}$ at 29.1 cm⁻¹ as the shear mode vibrating along the $a$-axis in $T'_{or}$-MoTe₂. We note that the DFT calculations as shown in Table 3 have an error of about 3 cm⁻¹ in determining the mode energies.

　　With an overall picture of the lattice vibrations as described above, we are now in a position to build an intuitive link between inversion symmetry breaking and Raman scattering of the two shear modes in the $T'$-MoTe₂ crystals. As we have discussed, this breaking of inversion symmetry is critically important for supporting Weyl fermions in $T'$-TMDC. We first note that $T'_{mo}$-MoTe₂, like the $T'_{or}$ phase, has two layers of MoTe₂ in its unit cell. The monoclinic crystal thus also supports two similar shear vibrations, one $m$-mode along the $b$-axis, and one $z$-mode along the $a$-axis, with energies of 15.3 and 29.3 cm⁻¹ respectively according to DFT calculations in Table 2. The reason why these modes evade Raman scattering measurements at room temperature in Fig.2(c) is closely linked to the inversion symmetry and the position of inversion centers. The presence of inversion symmetry in $T'_{mo}$-MoTe₂ dictates that all the zone-center lattice vibrations have either even or odd parity, and the odd ones are Raman inactive due to selection rules. With the inversion centers located inside the MoTe₂ atomic layers (Fig.3(b)), we observe that the two shear modes calculated to be at 15.3 and 29.3 cm⁻¹ in Table 2 are odd under the inversion operation. This explains why at RT the $T''$-MoTe$_2$ Raman spectra do not show shear modes in Fig.2(c). It is interesting to note that bulk $H$-MoTe$_2$ displays its shear mode at 27.5 cm$^{-1}$ in Fig.2(b), in spite of being inversion symmetric. This is because for $H$-TMDC the inversion centers are located in-between the atomic layers (Fig.3(d)), making the centrosymmetric crystal's doubly degenerate shear modes even under the inversion operation and Raman active.

Since the shear modes in $T'_{mo}$-MoTe$_2$ have odd parity, the emergence of shear Raman intensity at low temperatures is an indication of cooling induced structural phase transition that breaks the inversion symmetry. Indeed as we have discussed and shown in Fig.3(c), $T'_{or}$-MoTe$_2$ is not centrosymmetric. The process of inversion symmetry breaking can be monitored by measuring the evolution of the Raman spectra as the temperature changes. Figure 4(a) shows typical evolution of $m$-mode Raman spectra with energies less than 150 cm$^{-1}$ when the sample is cooled down from RT to 78 K and then warmed back up to RT. The Raman bands between 70 and 130 cm$^{-1}$ appear at all temperatures, with slight changes in peak position and linewidth due to cooling or warming. The two new $m$-modes at 12.6 and 130.8 cm$^{-1}$ which only occur in the $T'_{or}$ phase are found to be sensitive to whether the temperature is going down or up; at 236 K, the peak intensity is much larger during warming than during cooling. This suggests that there is significant hysteresis in the $T'_{mo} \rightarrow T'_{or} \rightarrow T'_{mo}$ phase transition. In Fig.4(b) we have plotted the temperature dependence of the Raman intensity for the shear mode $m_{or}^{12.6}$ during cooling and warming. The originally missing $m_{or}^{12.6}$ persists up to RT when we warm up from LT and we had to heat the crystal up to 339K to make it completely disappear. At low temperatures the Raman intensities of $m_{or}^{12.6}$ tend to stabilize below 200 K and become independent of cooling or warming. This indicates that the crystal is stabilized in pure $T'_{or}$ phase, without any admixture from the $T'_{mo}$ at low temperatures, making it suitable for probing type II Weyl physics. The shear mode has low energy and requires relatively specialized Raman system to perform the measurement; however we see that the mode at 130.8 cm$^{-1}$ (highlighted by an asterisk in Fig.4(a)), which should be easily accessible to most Raman setups, displays behavior similar to the shear $m_{or}^{12.6}$: it appears only in the $T'_{or}$ phase and has hysteresis in concert with the $m_{or}^{12.6}$. We thus conclude that the $m_{or}^{130.8}$ mode provides the most convenient signature for monitoring the inversion symmetry breaking and the phase transition to the $T'_{or}$ structure. One could, in principle, use instead $z_{or}^{29.1}$ and $z_{or}^{186.8}$ in Fig.2(c) and (e); however we have found that these peaks are much weaker than $m_{or}^{12.6}$ and $m_{or}^{130.8}$.

In conclusion, we have probed with Raman scattering the inversion symmetry and the crystal phase transition of $T'$-MoTe$_2$. Our investigation provides a generic approach for analyzing and detecting the lattice $m$-mode and $z$-mode vibrations. The two new shear modes that we observed and systematically analyzed were found to be directly linked to inversion symmetry breaking in the $T'_{mo}$-$T'_{or}$ structural phase transition in the crystal. The two concomitant high energy modes, especially the $m_{or}^{130.8}$ mode, provide a very convenient Raman fingerprint for the $T'_{or}$ phase that has raised much recent interest for studying type II Weyl fermions. We further anticipate that the cooling-driven inversion-symmetry breaking might also be probed by second harmonic generation[34,35]. Finally, the thermally-driven stacking changes could also occur in atomically-thin $T'$-MoTe$_2$, raising interesting questions regarding stacking-dependent vibrational, optical and electronic properties, which are known to display rich physics in other 2D semimetals such as graphene[36–38].

**Methods**

**Crystal growth**. The $T'$-MoTe$_2$ crystal used in this work is grown via chemical vapor transport method using bromine as the transport agent, as illustrated in Fig.1(a). Mo, Te, and TeBr$_4$ powders are placed in a fused silica tube, 18 mm in diameter and 300 mm in length. The purity of the source materials are Mo 99.9 %, Te 99.997 %, and TeBr$_4$ 99.999 % (Sigma Aldrich). Total Mo and Te are kept in a stoichiometric 2:1 ratio with sufficient TeBr$_4$ to achieve a Br density of 3 mg/cm$^3$. The tube is pump-purged with ultra-high purity argon gas and sealed at low pressure prior to growth. A three-zone tube furnace is used to provide a high temperature reaction zone and a low temperature growth zone. The reaction and growth zones were kept at 1000 °C and 900 °C respectively for 100 hours. At the end of the growth the crystal is thermally quenched in a water bath to keep the crystal from transitioning into the hexagonal ($H$) phase.

**DFT calculation**. We use plane-wave density functional theory (DFT) as implemented in the Vienna Ab Initio Simulation Package (VASP)[26] to calculate the 36 phonon branch dispersions of $T'_{mo}$ and $T'_{or}$ MoTe$_2$. The projector-augmented wave (PAW) method[39,40] was employed to represent core and valence electrons, the valence configurations of Mo and Te being 4p$^6$4d$^5$5s$^1$ and 5s$^2$5p$^4$ respectively. From convergence tests, a plane-wave cutoff of 325 eV was chosen in conjunction with a $\Gamma$-centered, $4 \times 8 \times 2$ k-point mesh for Brillouin zone sampling. For relaxation of the primitive cell, electronic wavefunctions were converged to within 10$^{-4}$ eV; cell vectors and atomic positions were optimized with a stress tolerance of 1 kbar and a force tolerance of 10$^{-2}$ eV/Å, respectively, followed by an additional relaxation of atomic positions alone with a force tolerance of 10$^{-3}$ eV/Å . As standard DFT functionals fail to describe interlayer van der Waals bonding correctly, we used the non-local optB86b van der Waals functional[27,28], which reproduces the equilibrium geometry of MoTe$_2$ accurately[29]. Following structural optimization of primitive cells, phonon dispersions were obtained within the harmonic approximation using the finite-displacement method in Phonopy[41]. A $4 \times 4 \times 1$ supercell with a $1 \times 2 \times 2$ k-point mesh was employed for the $T'_{mo}$ phase, whereas a larger $4 \times 6 \times 1$ supercell was required for the $T'_{or}$ phase with a $1 \times 1 \times 2$ k-point mesh. Electronic wavefuncions were converged with a tighter cutoff of $10^{-6}$ eV in the force constant calculations. Tables 2 and 3 show the results of DFT calculation, including first Brillouin zone, phonon dispersion, character table for corresponding symmetry, zone center normal modes with their calculated energies as well as vibrational symmetry representations.

## Author contributions

S.-Y.C. and J.Y. conceived the optical experiment. T.G. conceived the CVT growth of bulk MoTe$_2$. S.-Y.C. performed the polarization and crystal orientation-resolved Raman scattering measurements. A.R. performed the DFT calculation. D.V. conceived the X-ray diffraction measurements. S.-Y.C., A.R. and J.Y. co-wrote the paper. All authors discussed the results, edited, commented and agreed on the manuscript.

## Notes

The authors declare no competing financial interests.

Recently we became aware of two related Raman studies of $T'$-MoTe$_2$.[42,43]

**Acknowledgements**

This work is supported by the University of Massachusetts Amherst and in part by the Armstrong Fund for Science and the National Science Foundation Center for Hierarchical Manufacturing (CMMI-1025020). A.R. gratefully acknowledges supercomputing support from the University of Massachusetts Amherst and the Massachusetts Green High Performance Computing Center.

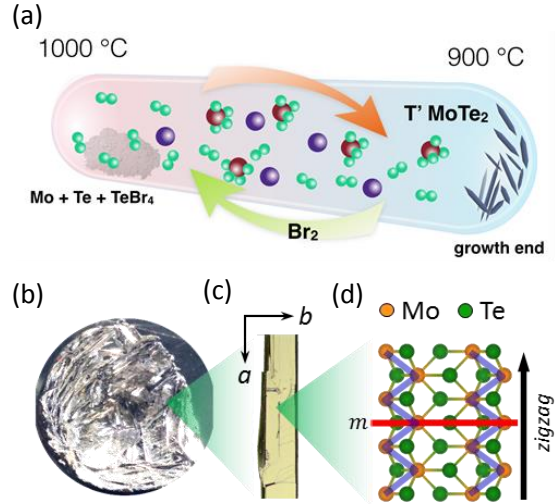

**Figure 1.** (a) Schematic of CVT growth of *T'*-MoTe$_2$. (b) Picture of a grown *T'*-MoTe$_2$ sample composed of many needle-like single crystals. (c) A zoomed-in optical image of a *T'*-MoTe$_2$ single crystal. The needle direction is along the *a*-axis. (d) Top view of atomic arrangement of a monolayer *T'*-MoTe$_2$. The *a*-axis points along the Mo-Mo zigzag chain (purple zigzags); and the *b*-axis lies in a mirror plane (thick red horizontal line) perpendicular to the zigzag chains. The crystal orientation relating (c) and (d) is confirmed by XRD.

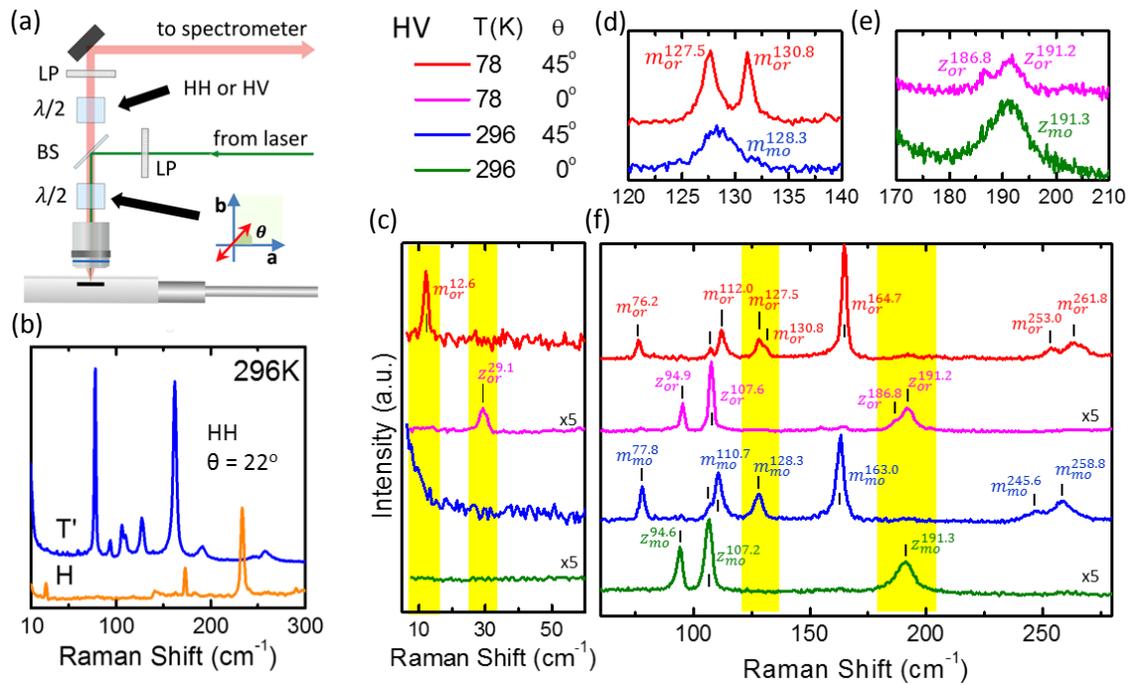

**Figure 2.** (a) Schematic of polarization-resolved and crystal-orientation-resolved Raman spectroscopy. In the inset, the angle $\theta$ is between the polarization of the incident laser light (red arrow) and the crystal $a$-axis. (b) Typical room temperature Raman spectra of $T'$-MoTe$_2$ and $H$-MoTe$_2$ in HH scattering configuration. For $T'$-MoTe$_2$, the angle $\theta = 22°$. (c-f) The Raman spectra of $T'$-MoTe$_2$ at 78 K and 296 K in HV with $\theta = 45°$ and $0°$. The yellow bands highlight the four emerging new modes at 78 K. Panels (d) and (e) show the zoomed-in spectra of the two new high energy modes taken with 3-fold higher spectral resolution.

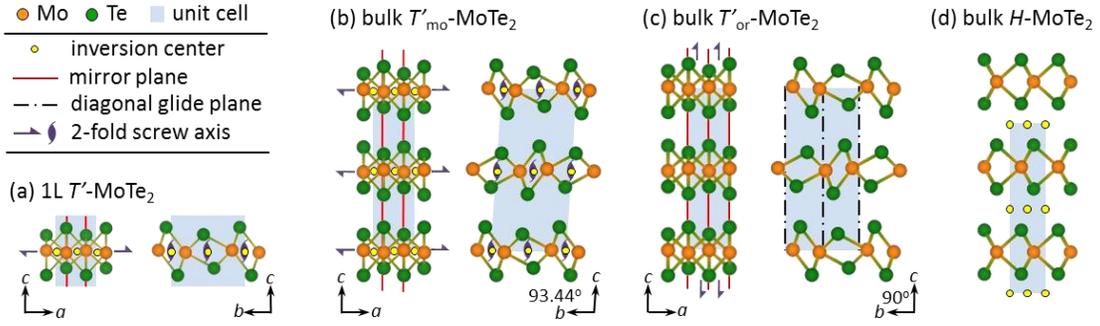

**Figure 3.** Crystal structure and the position of symmetry operators of (a) 1L $T'$-MoTe$_2$, (b) bulk $T'_{mo}$-MoTe$_2$, (c) bulk $T'_{or}$-MoTe$_2$ and (d) bulk $H$-MoTe$_2$. The notations for symmetry operations are consistent with International Tables for Crystallography[23]. The shaded areas indicate the unit cell in the corresponding phases. The inversion centers are located inside the atomic layers for $T'$-MoTe$_2$, and in-between the atomic layers for bulk $H$-MoTe$_2$.

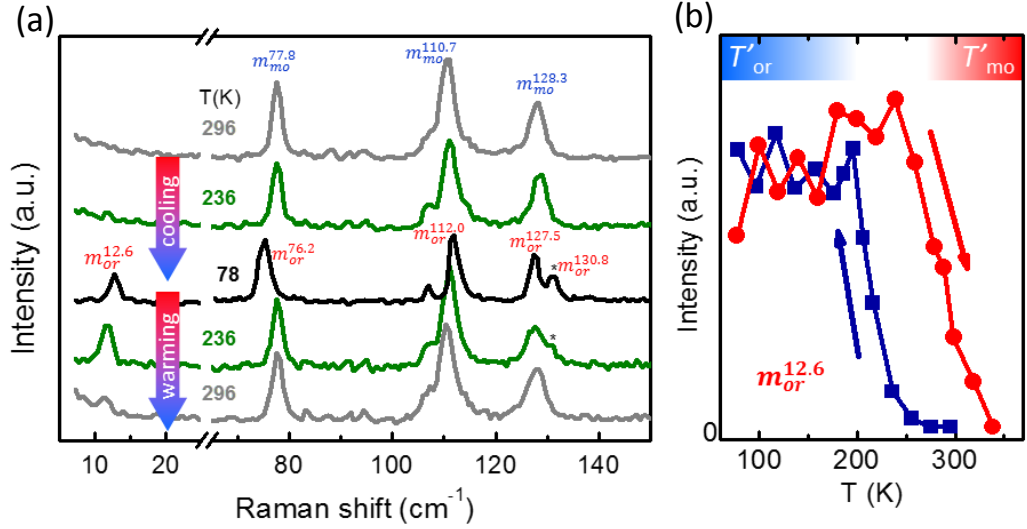

**Figure 4.** The $T'$-MoTe$_2$ $m$-mode Raman spectra with energy less than 150 cm$^{-1}$ under different thermal cycles. The Raman spectra collected here are dispersed by a single grating. Two modes $m_{or}^{12.6}$ and $m_{or}^{130.8}$ emerge when the sample cools down from 296 K to 78 K and persist during warming up to 296 K. (b) Temperature dependent intensity of $m_{or}^{12.6}$ mode during cooling (dark blue) and warming (red). The hysteresis means that $T'_{mo}$ and $T'_{or}$ phases can coexist in certain temperature range. For temperatures lower than 200 K, the intensity overlaps for cooling and warming, indicating a complete phase transition from $T'_{mo}$ to $T'_{or}$.

**Table 1.** Angular dependence (with respect to the *a*-axis) of Raman intensity of $m_{or}$, $z_{or}$, $m_{mo}$ and $z_{mo}$ modes. The polarization configurations (HH or HV) and the mode energies are noted in each panel. The solid curves are fits using Equations (1) and (2) in the text.

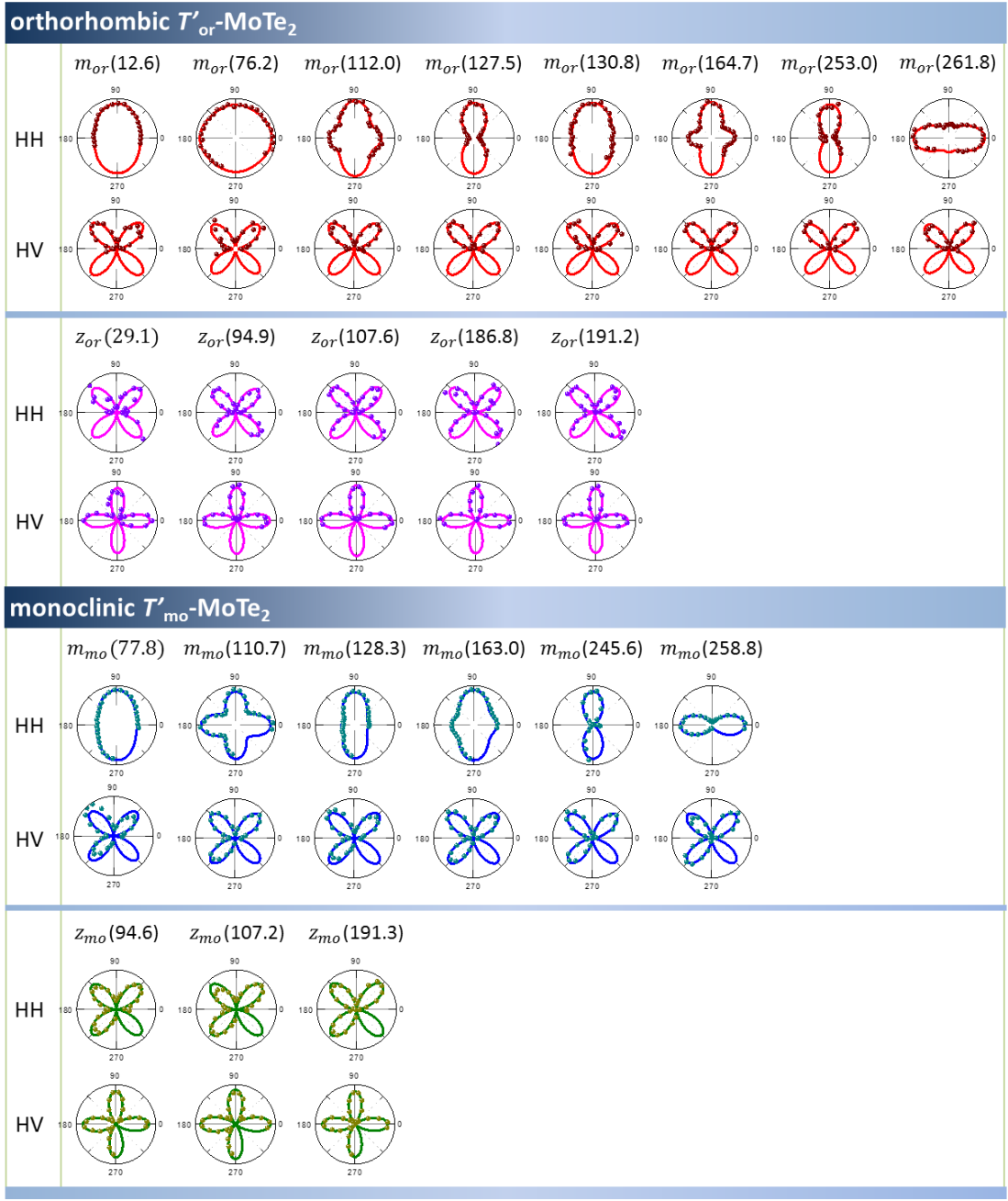

**Table 2.** DFT calculation results of $T'_{mo}$-MoTe$_2$, including first Brillouin zone, phonon dispersion, character table for $C_{2h}$ symmetry group, and the schematics of zone center normal modes with their calculated energies as well as the vibration symmetry representations.

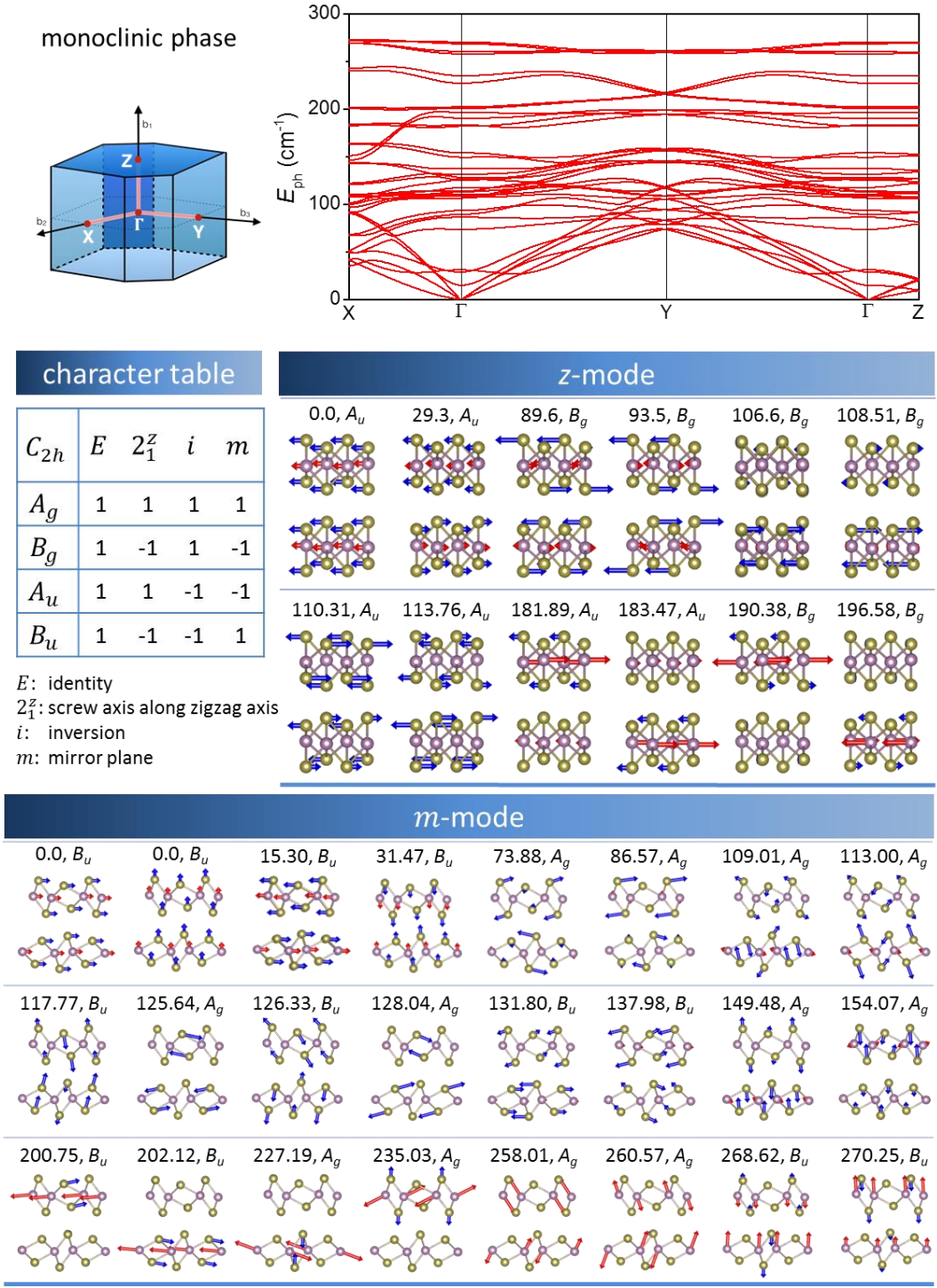

**Table 3.** DFT calculation results of $T'_{or}$-MoTe$_2$, including first Brillouin zone, phonon dispersion, character table for $C_{2v}$ symmetry group, and the schematics of zone center normal modes with their calculated energies as well as the vibration symmetry representations. The four new modes, $z_{or}^{29.1}$, $z_{or}^{186.8}$, $m_{or}^{12.6}$ and $m_{or}^{130.8}$ in Fig. 2(c)-(f) activated by the phase transition, are highlighted in yellow. The difference between theory and experiment in mode energy is around 3 cm$^{-1}$.

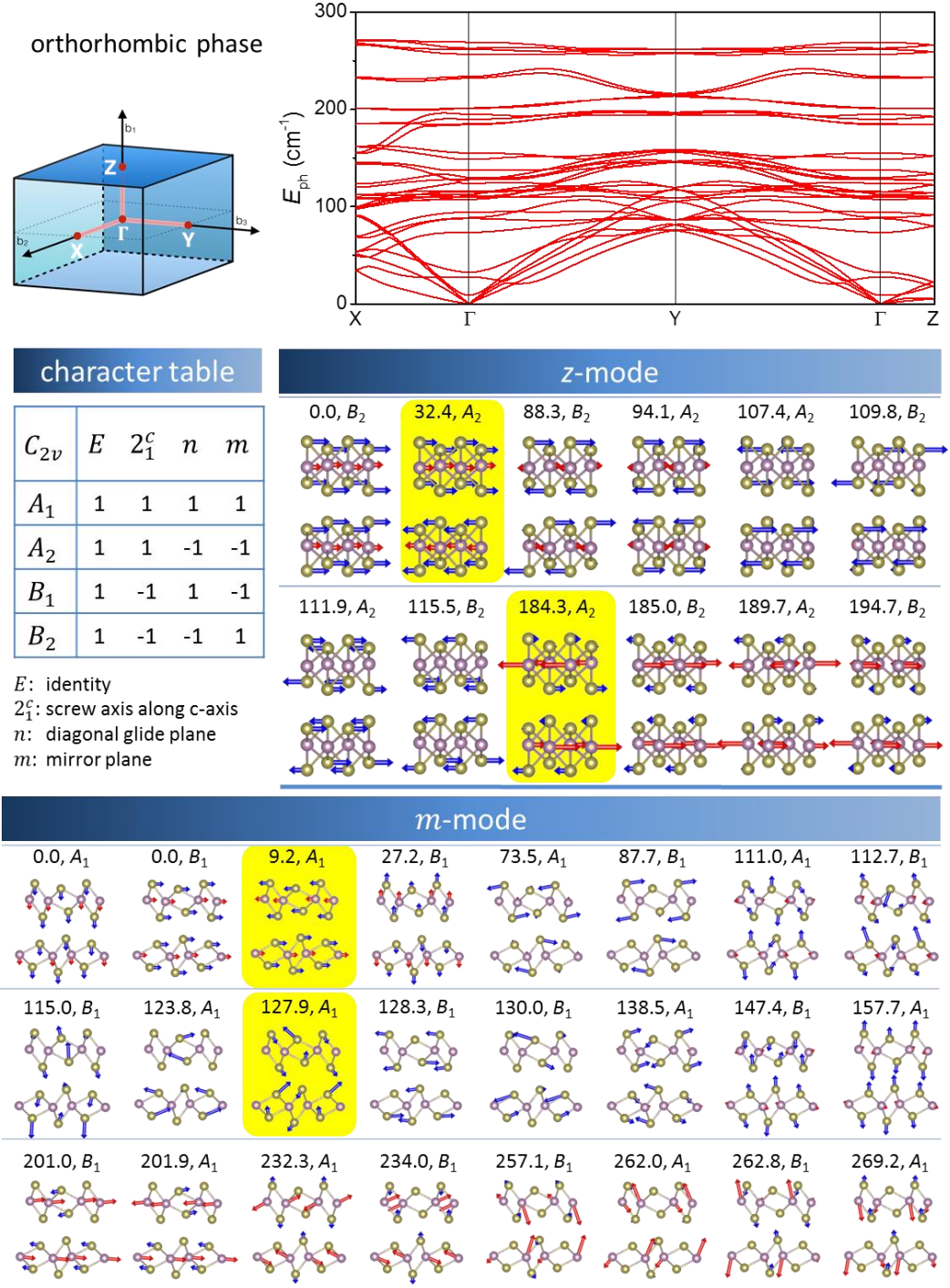

**TOC**

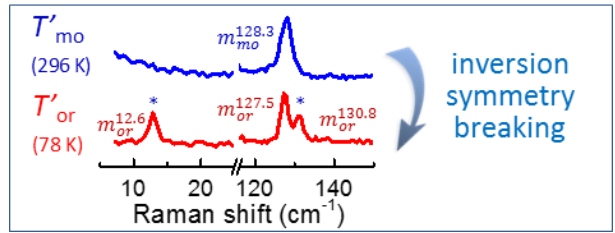